\documentclass{article}
\usepackage{amsmath,graphics,amsfonts,amsthm,amssymb,color}
  \usepackage{graphics}
   \usepackage{epsfig}
    \usepackage[colorlinks=true]{hyperref}
      \usepackage{cite}
       \usepackage{subfigure}
        \usepackage{url}

 \hypersetup{urlcolor=blue, citecolor=red}

\newcommand{\Rn}{\mathbb{R}^{n}}

\newcommand{\ao}{\alpha_{1}}
\newcommand{\at}{\alpha_{2}}
\newcommand{\bo}{\beta_{1}}
\newcommand{\bt}{\beta_{2}}

\title{Towards a New Spatial Representation of Bone Remodeling }

\begin{document}
\maketitle

%% Enter the first author's name and address:
\centerline{\scshape Jason M. Graham }
\medskip
{\footnotesize
 %% please put the address of the first author
 \centerline{Department of Mathematics Program in Applied Mathematical and Computational Sciences}
   \centerline{University of Iowa, Iowa City, IA 52242-1419, USA}
} %% Do not forget to end the {\footnotesize by the sign }

\medskip

\centerline{\scshape Bruce P. Ayati}
\medskip
{\footnotesize
 %% please put the address of the second author
 \centerline{Department of Mathematics Program in Applied Mathematical and Computational Sciences}
   \centerline{University of Iowa, Iowa City, IA 52242-1419, USA}
} %

\centerline{\scshape Prem S. Ramakrishnan}
\medskip
{\footnotesize
 %% please put the address of the second author
 \centerline{Department of Orthopaedics and Rehabilitation, University of Iowa Hospitals and Clinics}
   \centerline{University of Iowa, Iowa City, IA 52242-1419, USA}
} %

\centerline{\scshape James A. Martin}
\medskip
{\footnotesize
 %% please put the address of the second author
 \centerline{Department of Orthopaedics and Rehabilitation, University of Iowa Hospitals and Clinics}
   \centerline{University of Iowa, Iowa City, IA 52242-1419, USA}
} %

\bigskip

%% The name of the associate editor
%\centerline{(Communicated by the associate editor name)}

\begin{abstract}
Irregular bone remodeling is associated with a number of bone diseases such as osteoporosis and multiple myeloma.
 Computational and mathematical modeling can aid in therapy and treatment as well as understanding fundamental biology. Different approaches to
 modeling give insight into different aspects of a phenomena so it is useful to have an arsenal of various computational and mathematical models.
 Here we develop a mathematical representation of bone remodeling that can effectively describe many aspects of the complicated geometries and spatial behavior observed.

  There is a sharp interface between bone and marrow regions. Also the surface of bone moves in and out, i.e. in the normal direction, due to remodeling. Based on these observations we employ the use of a level-set function to represent the spatial behavior of remodeling. We elaborate on a temporal model for osteoclast and osteoblast population dynamics to determine the change in bone mass which influences how the interface between bone and marrow changes.

  We exhibit simulations based on our computational model that show the motion of the interface between bone and marrow as a consequence of bone remodeling. The simulations show that it is possible to capture spatial behavior of bone remodeling in complicated geometries as they occur \emph{in vitro} and \emph{in vivo}.

  By employing the level set approach it is possible to develop computational and mathematical representations of the spatial
behavior of bone remodeling. By including in this formalism further details, such as more complex cytokine interactions and accurate parameter values, it is possible to obtain simulations of phenomena related to bone remodeling with spatial behavior much as \emph{in vitro} and \emph{in vivo}. This makes it possible to perform \emph{in silica} experiments more closely resembling experimental observations.
\end{abstract}

\section*{Introduction}

Bones of the mature skeleton undergo constant remodeling through resorption and replacement of old matrix.  The repair of micro-cracks caused by normal mechanical stresses entails stimulation of local remodeling by tissue-intrinsic factors \cite{parfitt1994,robling2006,pobb}. Bone remodeling also plays a role in systemic $\mathrm{Ca}^{++}$ homoeostasis and is globally regulated by humoral factors (e.g. parathyroid hormone). Because the major mechanisms regulating
 bone metabolism are well-established and the links between abnormal metabolism and bone diseases are well-documented, the remodeling process is a potentially fertile ground for mathematical and computational analyses. Computer models that simulate the complexly regulated biology of bone remodeling can be used to rapidly and inexpensively screen drugs to estimate their disease-modifying activity. By minimizing the need for much more lengthy and expensive \emph{in vivo} testing, such models can greatly accelerate the development of novel treatments for bone diseases.

  Bone remodeling occurs simultaneously at various locations. However, remodeling from one site to another is not necessarily synchronous.
  Remodeling is localized to a particular remodeling site with teams of cells acting together as an individual structure commonly referred to as
  a basic multicellular unit (BMU) \cite{parfitt1994,robling2006,pobb,raisz1999}.  Remodeling is driven locally by signals from osteocytes, which secrete chemotactic factors in response to mechanically induced microfractures and in response to bone strain \cite{parfitt1994,robling2006,pobb}. With the release of factors by osteocytes stromal stem cells are activated and begin to produce factors, such as macrophage colony-stimulating factor (M-CSF), that promote osteoclastogenesis \cite{parfitt1994,pobb}. Additional stromal cells are created and differentiate into pre-osteoblasts. These cells begin producing factors such as receptor activator of nuclear factor-B ligand (RANKL), which binds to receptors present in pre-osteoclasts and inhibits apoptosis \cite{parfitt1994,pobb}.
  This is a case of osteoblast-derived paracrine signaling and is featured in many of the mathematical  treatments of remodeling discussed below. In the presence of the factors M-CSF and RANKL pre-osteoclasts fuse to become mature osteoclasts. Osteoclasts bind to bone surfaces and resorb the matrix by secreting acids and matrix proteases. Mature osteoclasts in turn produce growth factors and other signaling molecules, an example of osteoclast-derived autocrine signaling. Resorption also releases growth factors such as IGF and TGF-$\beta$ that act on cells of the osteoblast lineage, an example of osteoclast-derived paracrine signaling \cite{parfitt1994,pobb}.

   When resorption is nearly complete pre-osteoblasts mature into osteoblasts and stop secreting RANKL. Instead they produce OPG, a decoy receptor that blocks RANKL binding to osteoclasts. As a result, osteoclasts cease resorptive activity and undergo apoptosis. During formation some osteoblasts become trapped in the matrix as osteocytes, while some undergo apoptosis \cite{parfitt1994,pobb}. This completes the description of a typical scenario for remodeling by a BMU in response to mechanical stress. This description is helpful in the development of the mathematical
  and computational models discussed below.

  A first step towards formulating a theoretical representation of bone remodeling is to incorporate the mechanisms that regulate the process.
 Here this is the autocrine and paracrine biochemical signaling amongst cells.
 Komarova et al \cite{komarova2003,komarova2005} develop a temporal model consisting of ordinary differential equations (ODEs) describing
the local dynamics of the remodeling process. This model uses a power law approximation to capture the interaction of
cell populations through autocrine and paracrine signaling. An alternative approach to that in \cite{komarova2003,komarova2005} also capturing the cytokine signaling regulating bone remodeling is given by Pivonka et al in \cite{pivonka2008}. The model presented in \cite{pivonka2008} incorporates signaling through Hill functions rather than power laws. Both approaches model the changes in osteoclast and osteoblast populations in response to RANKL/OPG signaling. Ayati et al \cite{ayati2010} and Ryser et al \cite{ryser1,komarova2010}
build on the model from \cite{komarova2003} for the local dynamics to incorporate some of the spatial effects of bone remodeling. In Ayati et al \cite{ayati2010},
in the context of multiple myeloma, diffusion of cells is added to incorporate some spatial effects of bone remodeling, while in \cite{komarova2010} Ryser et al consider chemotaxis of osteoclasts in the presence of remodeling promoting factors.

In this article we employ a geometric method developed by Osher and Sethian \cite{osher1988} to present another approach to
capturing spatial effects of bone remodeling. We represent, via the so called level-set equation, the evolution of the interface between bone
and marrow regions in response to remodeling. The interface is described geometrically as a closed curve in the plane. The interior of the curve models region of bone while the exterior to the curve represents the marrow region. A level-set function, which is determined by solving the level-set equation, provides a convenient way to represent a geometric object such as a curve. Through this geometric approach it is possible to consider effects of remodeling for complex geometries and spatial scales not easily represented with previous spatial models. This is particularly relevant for representing the remodeling of cancellous or trabecular bone tissue.  By interpreting the change in total bone mass, at each point in space, during remodeling
as the speed of the interface motion (normal, i.e. perpendicular, to the interface)\footnote{More precisely, at each point of the interface one can put a tangent line. By normal we mean perpendicular to this tangent line.} we obtain a novel representation of the spatial behavior of bone remodeling. We note that while in many other applications of this geometric approach, the motion of the interface is determined by some aspect of the local geometry such as curvature, this is not the case for the application we give here. The motion of the interface is a result of the behavior of the cell populations at each point in space, in other words, the changes in geometry are directly determined by biological phenomena alone.  We find
that our modeling and simulation efforts are representative of observations in animal models. Incorporating pathological behavior associated with abnormal bone remodeling in future work is straightforward and of clinical interest.

\section*{Models}

The complex nature of the interactions of osteoclasts and osteoblasts and the chemical signaling that regulates them makes a
  detailed theoretical and experimental study difficult. Mathematical modeling may be used to tie existing knowledge about a system
  with new ideas. Simulations based on mathematical models can provide a useful description of observed or hypothesized results otherwise difficult with ordinary language. In this section we present and discuss some mathematical models of bone remodeling with a focus on the ones that have influenced the development of our approach. The models presented here describe the change in osteoclast and osteoblast populations throughout the remodeling process. Throughout we denote by $C$ the osteoclast population and $B$ the osteoblast population.

\begin{figure}[!ht]
  \centering
  \includegraphics[width=60mm,height=60mm]{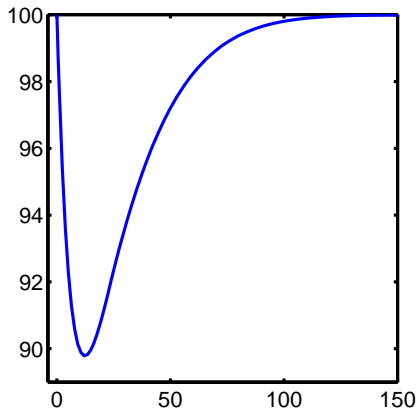}
  \caption{{\bf{Change in bone mass according to the model from \cite{komarova2003}.}} This figure was obtained by solving equations (\ref{eq:clasts1}),(\ref{eq:blasts1}), and (\ref{eq:bonemass1}). Here the horizontal axis is time in days and the vertical is percentage of initial (100\%) bone mass. }\label{komFig}
\end{figure}

 The model which forms the foundation for all that follows is the temporal model of Komarova et al \cite{komarova2003} that describes the local dynamics of bone remodeling at a single remodeling site. The dynamics of the cell populations is determined by the interaction of cells through autocrine and paracrine signaling. In \cite{komarova2003} the authors employ a power law approximation to describe the overall effects of the biochemical factors involved in signaling. This type of modeling, while simple to describe, is effective in capturing the nonlinear behavior of biochemical signaling. The exponents in the power law approximation describe the effectiveness of the autocrine and paracrine signaling on each cell type. From \cite{komarova2003}
 the dynamics of the cell populations are given by
 \begin{subequations}
 \begin{align}
   \frac{dC}{dt} & = \ao C^{g_{11}}B^{g_{21}} - \bo C ,\label{eq:clasts1} \\
   \frac{dB}{dt} & = \at C^{g_{12}}B^{g_{22}} - \bt B. \label{eq:blasts1}
 \end{align}
 Here $\ao$ and $\at$ are constants that reflect the rate of cell production while while $\bo$ and $\bt$ are constants reflecting the rate of cell removal of osteoclasts and osteoblasts respectively. As mentioned above the exponents $g_{ij}$ describe the effectiveness of the different types of signaling on each cell type with $g_{11}$ describing the effectiveness of osteoclast derived autocrine signaling, $g_{21}$ the effectiveness of
 osteoblast derived paracrine signaling, $g_{12}$ the effectiveness of osteoclast derived paracrine signaling and $g_{22}$ the effectiveness of
 osteoblast derived autocrine signaling. The authors use this model to simulate the behavior of remodeling and find that the value of the parameter $g_{11}$ for the effectiveness of osteoclast derived autocrine signaling most strongly effects the mode of remodeling behavior.

 The authors also calculate the change in total bone mass due to remodeling via the equation
 \begin{align} \frac{dz}{dt} &= -k_{1}\max\{0,C-\bar{C}\} + k_{2}\max\{0,B - \bar{B}\}. \label{eq:bonemass1} \end{align}
 \label{eq:sysone}
 \end{subequations}
Here $z$ is the bone mass $k_{1},k_{2}$ are constants describing the activity of bone resorption and formation respectively. The constants
 $\bar{C}$ and $\bar{B}$ denote the stable steady state values of osteoclasts and osteoblasts respectively and here are given by
 \begin{subequations}
 \begin{align}
   \bar{C} & = \left(\frac{\bo}{\ao} \right)^{\frac{1-g_{22}}{\gamma}}\left( \frac{\bt}{\at}\right)^{\frac{g_{21}}{\gamma}} \label{eq:cbar} ,\\
   \bar{B} & = \left(\frac{\bo}{\ao} \right)^{\frac{g_{12}}{\gamma}}\left( \frac{\bt}{\at}\right)^{\frac{1-g_{11}}{\gamma}},  \label{eq:bbar}\\
   \gamma & = g_{12}g_{21} - (1-g_{11})(1-g_{22}).
 \end{align}
 \label{eq:systwo}
 \end{subequations}
 In the level-set representation that follows we take the change in bone mass
 as giving the speed of motion of the interface in the normal direction. We note that in the work of Pivonka et al \cite{pivonka2008} there is an equation for the change in bone volume similar to (\ref{eq:bonemass1}). Thus the equations from \cite{pivonka2008} can be used in the following in place of (\ref{eq:clasts1}) and (\ref{eq:blasts1}). However, for computational simplicity we use the power law approximation since the resulting interface dynamics should be the same, for the same $z$ field, regardless which model is used for the cell populations is used to obtain such a $z$ field. For example, in our case we can obtain the same $z$ field from a system with and without explicit chemical concentrations. For convenience we reproduce (see figure \ref{komFig}) the change in bone mass computed in \cite{komarova2003} as a percentage of initial bone mass using the system (\ref{eq:clasts1}),(\ref{eq:blasts1}), and (\ref{eq:bonemass1}). Figure \ref{komFig} can then be used for qualitatively comparison with the results in this paper.

 In \cite{ayati2010} the authors give an implicit spatial representation of bone remodeling
 by interpreting the bone mass equation as reflecting bone thickness. We use this interpretation
 with the level set method discussed below to give a first
 study of the dynamic behavior of the bone marrow interface due to remodeling. This assumes an unrealistic degree
 of homogeneity but is an instructive first step in developing the level set approach. The authors in \cite{ayati2010} go on to
 incorporate explicitly spatial effects through the addition of linear diffusion terms to the system (\ref{eq:sysone}). This is all done in the context of multiple myeloma bone disease since the goal of their work was to show how perturbation
 of the mathematical models for bone remodeling can exhibit pathological behavior as is seen in association with metabolic bone diseases. The modeling approach developed below can also be used to study how pathological behavior can arise from perturbations representing irregular remodeling.

 In \cite{komarova2010} the authors develop a spatial representation of bone remodeling by modifying the system (\ref{eq:sysone}) to include motility of osteoclasts and explicit representation of the RANKL and OPG fields. The model is particularly successful in capturing the motion and steering of the BMU and simulations in one and two dimensions show the distinct features of the cutting cone of osteoclast cells. Their results are obtained by including the chemotaxis of osteoclasts towards increasing concentrations of RANKL and explicitly the dynamics of the RANKL/OPG pathway. The equations for RANKL and OPG concentrations include porous flow of the chemicals, activation of osteclastogenesis by RANKL and binding of RANKL/OPG. The two dimensional simulations based on the model elegantly show steering, as a chemotactic response, of a remodeling BMU along a microfracture .

 Also of note are the mathematical models of Martin and Buckland-Wright \cite{buckwright2004,buckwright2005}. The model in \cite{buckwright2004} simulates erosion depth due to resorption by modeling the changes in cellular activity due to cytokine signaling. This is done using the equations for enzyme kinetics of Michaelis and Menten. The model in \cite{buckwright2004} differs from the mathematical models in \cite{komarova2003,komarova2005,pivonka2008,ayati2010,komarova2010} in that it does not describe changes cell populations while the model in \cite{buckwright2005} does include changes in the osteoblast population. The models found in \cite{buckwright2004,buckwright2005} provide another useful method for studying how changes in signaling are manifested as physical changes in the bone. The recent review paper of Pivonka and Komarova \cite{pivkom2010} gives a nice overview mathematical modeling in bone biology.

 In the models discussed so far bone remodeling is regulated by biochemical signaling alone. Mechanical signaling also plays an important role in the regulation of cells involved in bone biology. There is a substantial body of work devoted to combining the biochemical and mechanical stimuli in mathematical models to study the effects that the two different types of stimuli have together \cite{Geris1,Geris2,Geris3,Geris4,Geris5,Peiffer}.

  For example in \cite{Geris1} the authors couple mechanical
 stimuli through finite element analysis with partial differential equation models describing spatio-temporal changes in cell populations. These changes are due to cell migration (chemotaxis, haptotaxis), proliferation and differentiation. They also include equations describing angiogenesis.
 Biochemical signaling is incorporated by including interactions of cell populations with growth factors. This all leads to a mathematical framework that is successful in simulating the combined effects of biochemical and mechanical regulatory mechanisms. An interesting result is the prediction of over-load induced nonunion formation. The modeling approach just discussed is well suited to the study of wound and fracture healing and issues such as angiogenesis that arise under such situations. This is the case for example with larger fractures in or on cortical bone, rather than microfractures. Here we leave the role of angiogenesis to future considerations and focus on the spatial manifestations of other features of the remodeling process.
 %With trabecular bone, angiogenesis plays a lesser role since there may already be sufficient vascularization. {\bf rephrase mention of angiogenesis %so to avoid complications} Moreover, the methods developed in this paper are well suited for the complex geometries associated with trabecular bone %so we put off the incorporation of mechanical stimuli and angiogenesis for future work.

%\section{Level Set Method}

  The geometric approach mentioned, was initially developed by Osher and Sethian in \cite{osher1988} to study geometry driven motion of an interface
  and is particularly well suited for problems in which there is a definite ``inside'' and ``outside'' such as in the modeling of
  soap bubbles. The method has since been further developed and extended, see for example \cite{osher2003,sethian1996,osher1991,zhao1996,zhao1997}, and has found many applications in various areas of applied mathematics and engineering \cite{osher2003,sethian1996,zhao1997}. Since the surface of bone forms a distinct, sharp interface the level-set function seems well suited for representing physical changes
  to the surface of bone due to remodeling.

  We present here some basic background on the level-set function approach specific to representing
  the spatial effects of bone remodeling in trabecular bone tissue. For a more in depth treatment and other applications see the basic texts by Osher and Fedkiw \cite{osher2003} and Sethian \cite{sethian1996}. As stated previously, we would like to describe the interface between bone and marrow regions as a closed\footnote{We note that the curve need not be closed to apply the level set method. We take a closed curve and in particular a circle to make clear what is ``inside'' and what is ``outside''.} curve in the plane.  The basic idea is to represent an interface, geometrically a curve, surface, etc. implicitly
  as some level set, typically the zero level set, of a higher dimensional object described by a so called level-set function. For example consider the function $\phi(x,y) := 1 - x^2 - y^2$. The zero level set of $\phi(x,y)$ is the set of all points $(x,y)$ in the plane satisfying
  \begin{equation}
    \phi(x,y) := 1 - x^2 - y^2 = 0,
  \end{equation}
  which describes a circle of radius 1 centered at the point $(0,0)$. Alternatively one could represent an interface locally by explicit functions, but this may require many different such function, or parametrically. However, the approach to representing geometry via a level-set function provides some advantages over these alternatives that make it well suited for applications.

  One advantage to representing an interface, in this case a curve, implicitly as above is that most of the geometric information one would like to have
  about the interface can be obtained from the level set function. For more details on this see the first chapter of the book by Osher and Fedkiw \cite{osher2003}. What really makes the level-set function appealing for applications such as the changes to bone surface due to remodeling, is that, given an initial interface represented by a specified level set function, there is a simple way to evolve the interface, by evolving the level set function in time.  More precisely, to represent the evolution of an interface we use a family of
  level-set functions $\phi(\bar{x},t)$ parameterized by time $t$ so that at each time $t$ the zero level set $\{\bar{x}:\ \phi(\bar{x},t) = 0\}$
  represents the interface at time $t$. Here $\bar{x} \in \Rn$  for some positive integer $n$, the case we are concerned with has $n=2$. Then
  following Osher and Sethian in \cite{osher1988} we can derive a differential equation for the motion of the interface in the normal
  direction due to normal velocity with speed given by a function $a(\bar{x},t)$. The equation is
  \begin{equation} \phi_{t}(\bar{x},t) = a(\bar{x},t)|\nabla \phi(\bar{x},t)| \label{eq:lse} \end{equation}
   and is known as the level-set equation. Here $\phi_{t}$ denotes the derivative with respect to time, $\nabla \phi(\bar{x},t)$ is the gradient of $\phi$ in the spatial variables and $|\bar{y}|$ is the length of a vector $\bar{y} \in \Rn$. In a sense the function $a(\bar{x},t)$ gives the
   rule telling what drives the motion of the interface and solving the level set equation describes what the interface looks like at any given time.
   Notice that here the level set equation does not contain a curvature term so that local geometry is not used to move the interface.

 The ODEs (\ref{eq:clasts1}) and (\ref{eq:blasts1}) effectively describe the local population dynamics of the primary cells involved in remodeling.
 Alternatively, these ODEs can represent a larger, but well-mixed spatial domain. We have discussed previous approaches to incorporating spatial heterogeneity into mathematical descriptions  of bone remodeling. Even with the success of these models it is valuable to move towards new and different spatial representations of bone remodeling. In that direction we employ the level-set function approach to develop such a representation.
 The interface between bone and marrow is distinct and sharp and we use this to motivate representing the system as one with a moving interface.
 We use the level-set equation discussed previously to describe the motion of this interface.

 Since the power law approximation of \cite{komarova2003} successfully captures the local dynamics of osteclast/osteoblast cell populations we elaborate (\ref{eq:clasts1}) (\ref{eq:blasts1}) by adding explicit space dependence and coupling  with the level-set equation to represent spatial manifestations of the remodeling process. In other words, we (locally) couple the biological aspects, i.e.\ population dynamics of remodeling cells, of bone bone remodeling with the change in geometry of the interface between bone and  marrow regions. This is done by assuming that the speed in the direction normal to the interface, $a(x,t)$ in (\ref{eq:lse}), is given by the change in bone mass. More precisely, the speed in the normal direction is given by a constant multiple of the change in bone mass,
 \begin{equation} a(x,t) = \nu(-k_{1}\max\{0,C(x,t)-\bar{C}\} + k_{2}\max\{0,B(x,t)-\bar{B}\}), \label{eq:normv} \end{equation}
 where $\nu$ is a constant. We note that here there is a (potentially) different population of each cell type at each location $x$ in space. This makes the speed $a(x,t)$ different at each point in space.  In \cite{komarova2010} chemotaxis plays the role of a steering mechanism, determining the spatial behavior of the osteoclast and osteoblast populations along an existing microfracture on the order of a few microns. In our model the spatial behavior, including steering of cells, is determined by the geometry via the level set method. This is appropriate for spatial scales on the order of a thousand microns or larger and is well suited for geometrical configurations common in trabecular bone. If necessary, slight modification of our approach allows for the inclusion of chemotaxis or other steering mechanisms.

\section*{Results and Discussion}

In what follows, we denote by $\Gamma_{t}=\{x:\ \phi(x,t) = 0\}$ the interface at time $t$, where $\phi$ is the level-set function. We solve the following system of equations to simulate the motion of the interface between bone and marrow due to bone remodeling,
 \begin{subequations}
 \begin{align}
   \frac{\partial C}{\partial t}(x,t) & = \ao C(x,t)^{g_{11}}B(x,t)^{g_{21}} - \bo C(x,t) \label{eq:lvm1} \\
   \frac{\partial B}{\partial t}(x,t) & = \at C(x,t)^{g_{12}}B(x,t)^{g_{22}} - \bt B(x,t), \label{eq:lvm2} \\
   \phi_{t}& = a(x,t)|\nabla \phi|, \label{eq:lvm3} \\
   C(x,t)& = \tilde{C}(x,t), &x\in \Gamma_{t} \label{eq:Cupdate} \\
   B(x,t)& = \tilde{B}(x,t), &x\in \Gamma_{t}\label{eq:Bupdate}
 \end{align}
 \label{eq:lvmsys}
 \end{subequations}
 where $a(x,t)$ is given by (\ref{eq:normv}), and $\tilde{C}(x,t)$ and $\tilde{B}(x,t)$ are the number of osteoclasts and osteoblasts respectively, recruited from the marrow to a point $x$ on interface at time $t$.
 Thus we have a system of equations where $C(x,t),B(x,t)$ depend on the interface, i.e.\ level set function $\phi(x,t)$, since at each time, the population at the interface $\Gamma_{t}$ is updated according to (\ref{eq:Cupdate}) and (\ref{eq:Bupdate}). The rapid motion of the cells from deeper within the marrow to a remodeling site is captured by the local recruitment terms (\ref{eq:Cupdate}) and (\ref{eq:Bupdate}). When remodeling is initiated the cells are recruited quickly and directly to the remodeling site. As a result, diffusion of these cells is not relevant for the purposes we are considering. The domain for this problem is a square. To mimic a large domain, for computational simplicity we assume periodic boundary conditions for this domain. The initial interface is described by the zero level set of an initial level-set function which we denote by $\phi_{0}(x,y)$. For this discussion we take the initial interface to be a circle
 but note that any other shape could be taken
 by prescribing an appropriate level-set function. It is even possible to use, as a level set function, images taken from experiments, see for example \cite{sumengen}. For the numerical simulations presented below we use parameter values as in \cite{komarova2003,ayati2010,ryser1}, these are
 \begin{align*}
   \bar{C}& = 1.06 &
   \bar{B}& = 212.13\\
   \ao & = 3.0&
   \bo & = 0.2\\
   \at & = 4.0&
   \bt & = .02\\
   g_{11} & = 0.5&
   g_{21} & = -0.5\\
   g_{12} & = 1.0&
   g_{22} & = 0.0\\
   k_{1} & = 0.24&
   k_{2} & = 0.0017
 \end{align*}

 %\begin{figure}
 % % Requires \usepackage{graphicx}
 % \centering
 % \includegraphics[width=60mm,height=60mm]{initialInterface.eps}
 % \caption{Unit square domain with periodic boundary conditions along with initial interface. Interior of the circle represents a region of bone %while the exterior of the circle represents surrounding marrow. The initial interface is described by the zero level set of the level set function %$\phi_{0}(x,y) = \sqrt{\left(x- \frac{1}{2}\right)^{2} + \left(y - \frac{1}{2}\right)^{2}} - \frac{1}{4}$. The system (\ref{eq:lvmsys}) along with %the bone mass equation (\ref{eq:bonemass1}) describe how this interface will evolve in response to bone remodeling. }\label{initialInterface}
%\end{figure}

 All of our simulations have been carried out using MATLAB. The implementation of our model required solving a coupled system of differential equations at each point in the domain. To carry this out we first solved the equations for the cell populations using implicit time stepping and a Krylov iterative method \cite{kelley1,kelley2} to solve the nonlinear system.  This provided the normal speed for the motion of the interface. Evolving the interface required solving the level-set equation. For this we used the level set toolbox for MATLAB \cite{michell1} developed by Ian Mitchell. This toolbox contains many useful tools for implementing the level-set numerical method in MATLAB.

\begin{figure}[!ht]
   \centering
   \subfigure[$t = 0$ days]{\includegraphics[height=1.6in]{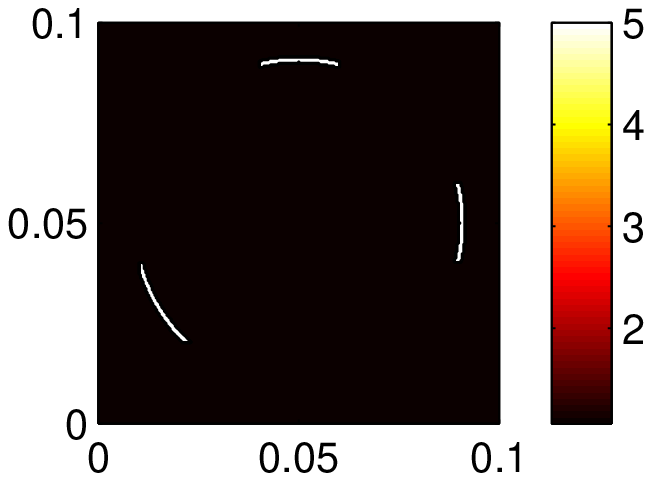}\label{oc-a}}
   \subfigure[ $t = 5$ days]{\includegraphics[height=1.6in]{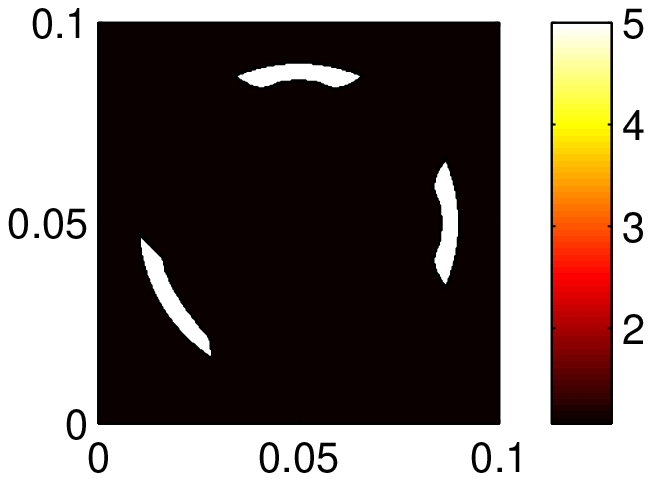}}
   \subfigure[ $t = 10$ days]{\includegraphics[height=1.6in]{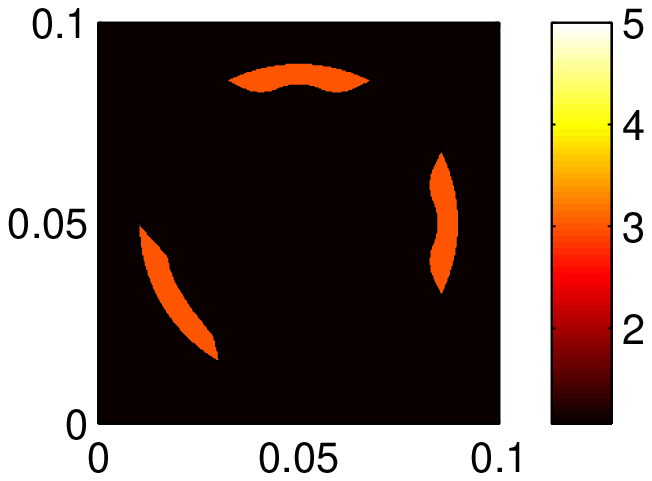}}\\
   \subfigure[ $t = 15$]{\includegraphics[height=1.6in]{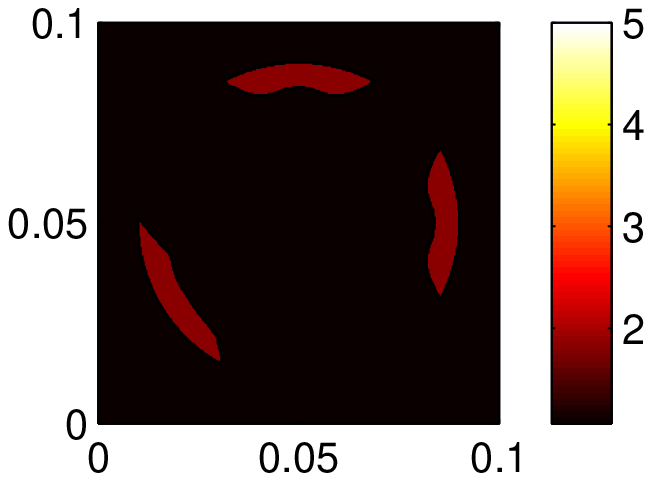}}
   \subfigure[ $t = 20$]{\includegraphics[height=1.6in]{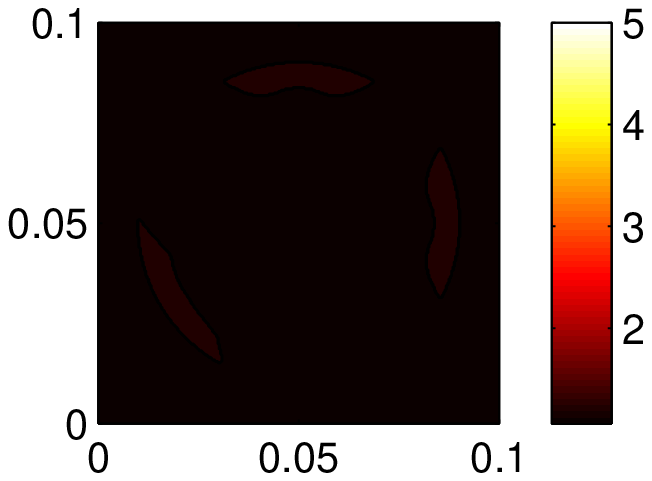}}
   \subfigure[$t = 25$ days]{\includegraphics[height=1.6in]{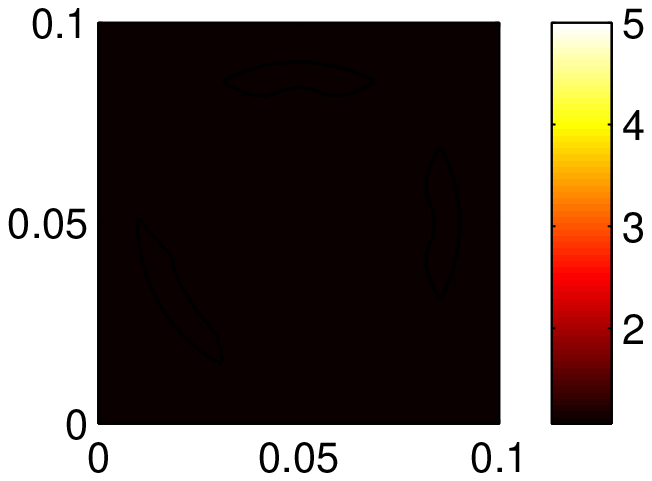}\label{oc-f}} \\
   \subfigure[ $t = 30$]{\includegraphics[height=1.6in]{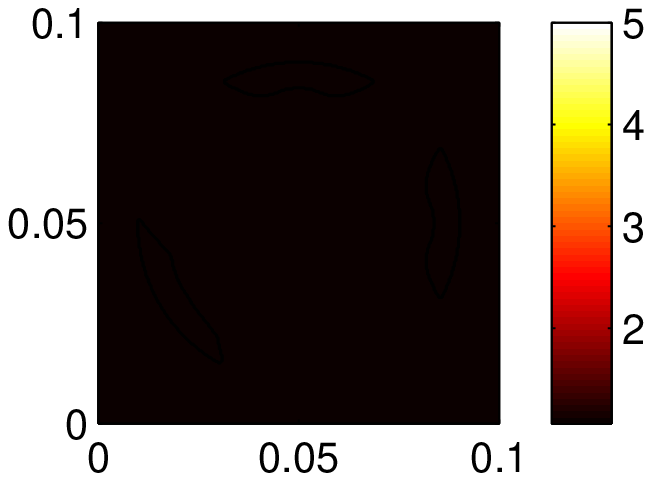}}
   \subfigure[ $t = 35$]{\includegraphics[height=1.6in]{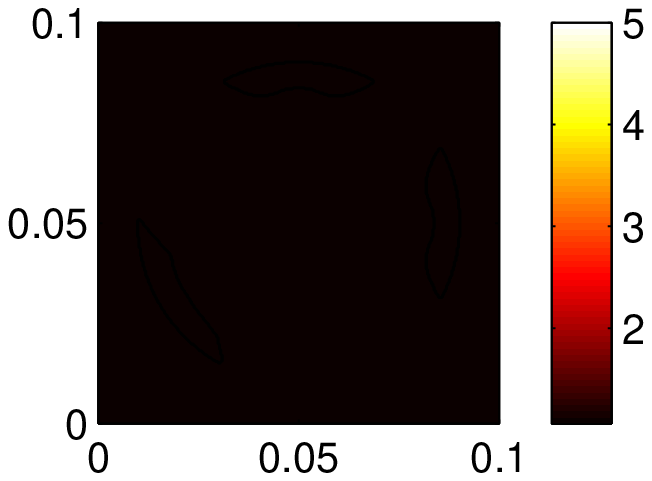}}
   \subfigure[$t = 150$ days]{\includegraphics[height=1.6in]{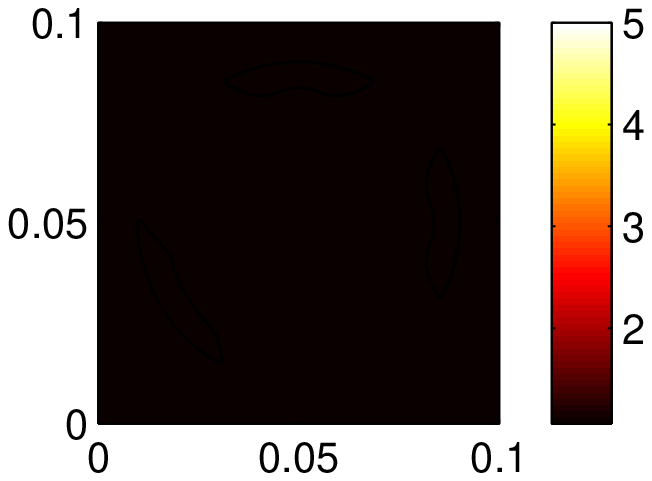}}
   \caption{{\bf{The population dynamics of osteoclast cells at three separate remodeling sites.}} Remodeling is initiated at three distinct sites by perturbing the osteoclast population away from steady state. The osteoclast population quickly returns to steady state. Bone is resorbed in the presence of an osteoclast population above steady state levels. }
   \label{osteoclasts}
 \end{figure}

\begin{figure}[!ht]
   \centering
   \subfigure[$t = 0$ days]{\includegraphics[height=1.6in]{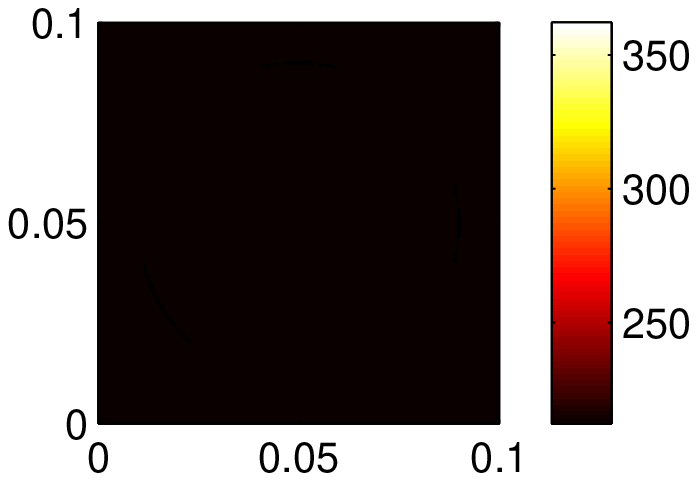}}
   \subfigure[ $t = 5$ days]{\includegraphics[height=1.6in]{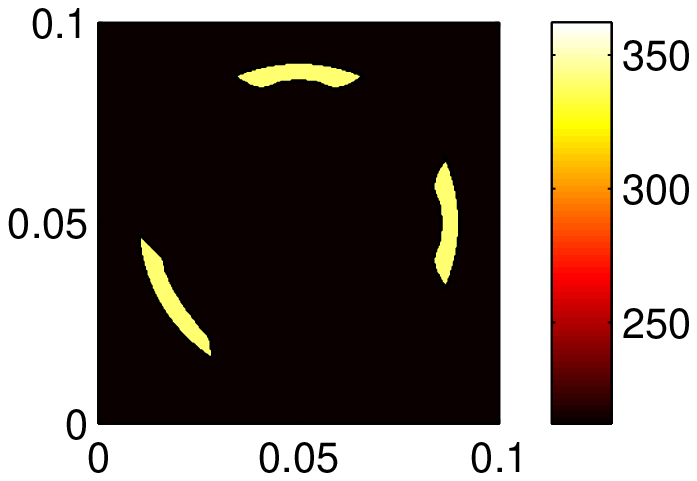}\label{ob-b}}
   \subfigure[ $t = 10$ days]{\includegraphics[height=1.6in]{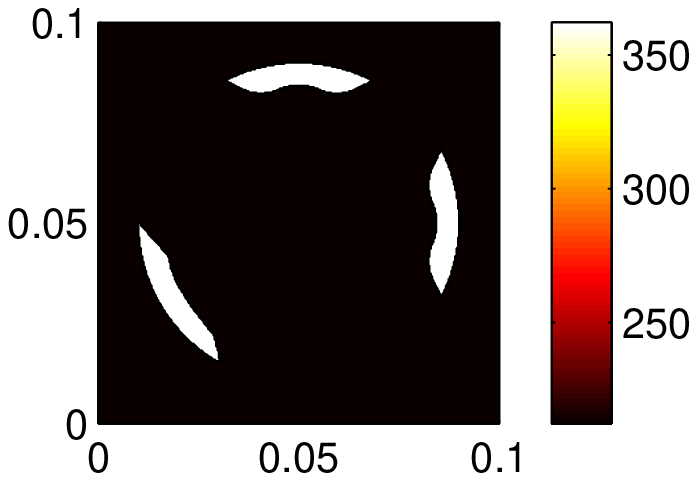}\label{ob-c}}\\
   \subfigure[ $t = 15$]{\includegraphics[height=1.6in]{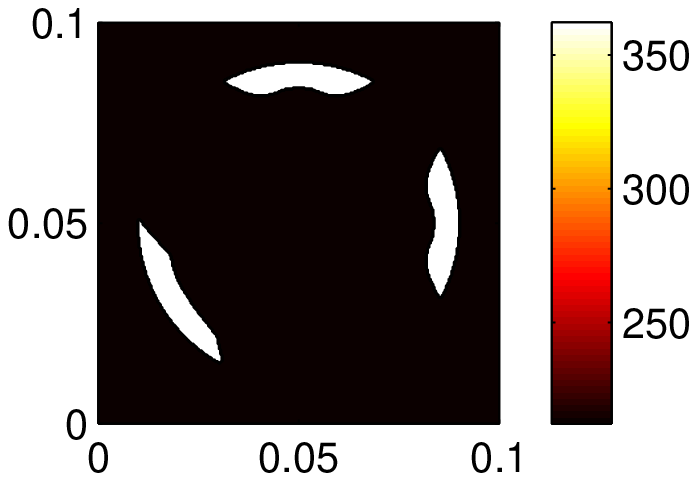}}
   \subfigure[ $t = 20$]{\includegraphics[height=1.6in]{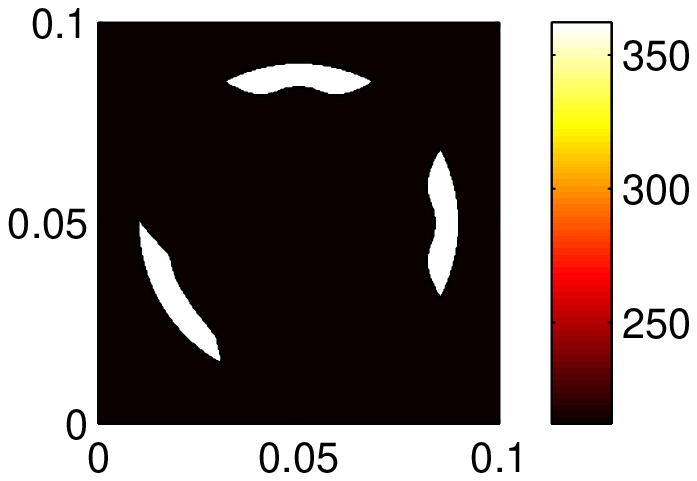}\label{ob-e}}
   \subfigure[$t = 25$ days]{\includegraphics[height=1.6in]{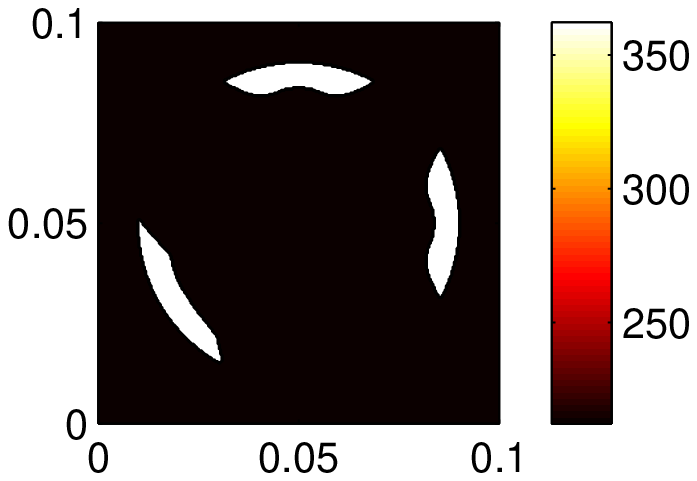}\label{ob-f}} \\
   \subfigure[ $t = 30$]{\includegraphics[height=1.6in]{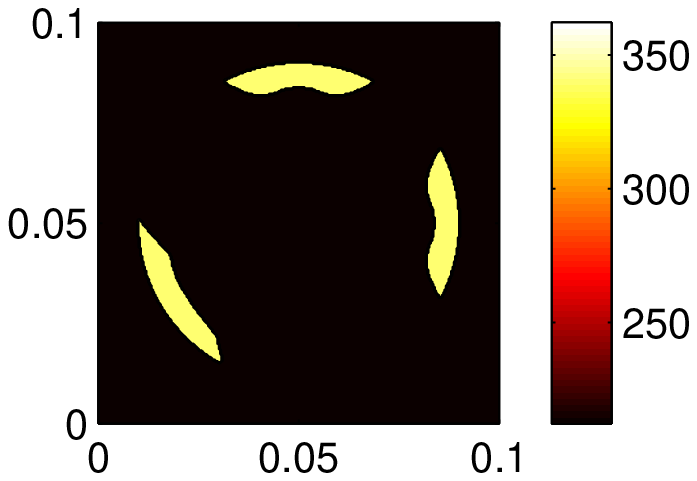}}
   \subfigure[ $t = 35$]{\includegraphics[height=1.6in]{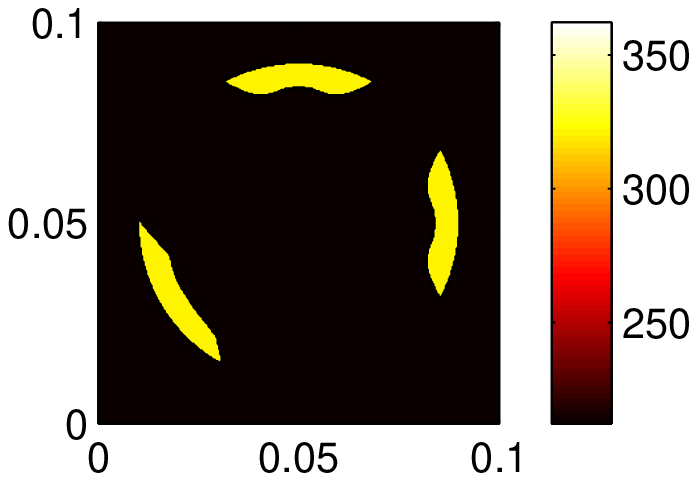}}
   \subfigure[$t = 150$ days]{\includegraphics[height=1.6in]{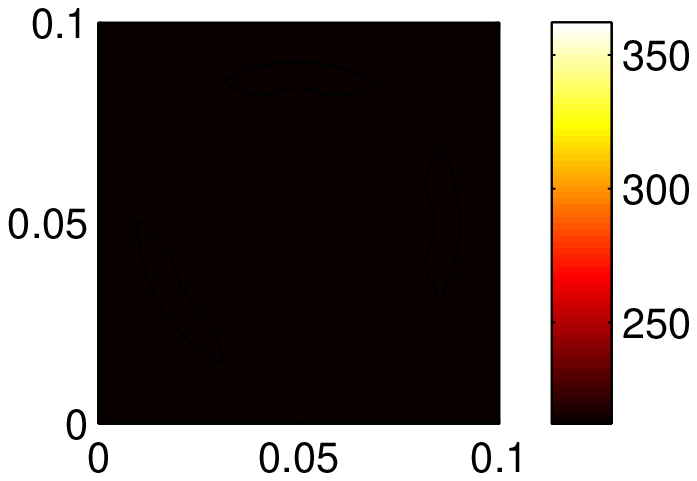}\label{ob-i}}
   \caption{{\bf{The population dynamics of osteoblast cells at three separate remodeling site.}} The osteoblast population starts at steady state, then increases in response to an increase in osteoclast population, and finally returns to steady in accord with (\ref{eq:clasts1}) and (\ref{eq:blasts1}).  Bone is formed in the presence of an osteoclast population above steady state levels.}   \label{osteoblasts}
 \end{figure}

\begin{figure}[!ht]
   \centering
   \subfigure[$t = 0$ days]{\includegraphics[height=1.5in]{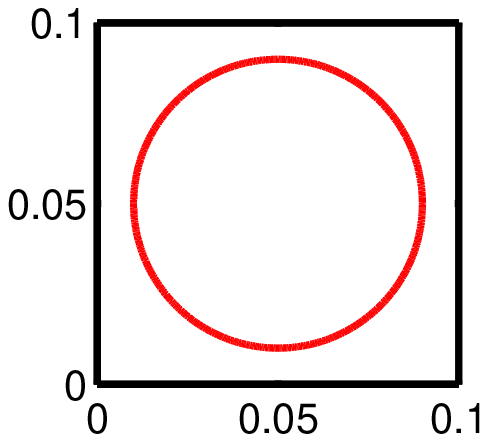}\label{if-a}}
   \subfigure[ $t = 5$ days]{\includegraphics[height=1.5in]{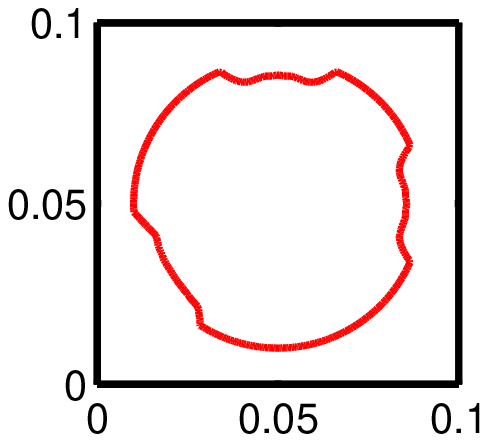}\label{if-b}}
   \subfigure[ $t = 10$ days]{\includegraphics[height=1.5in]{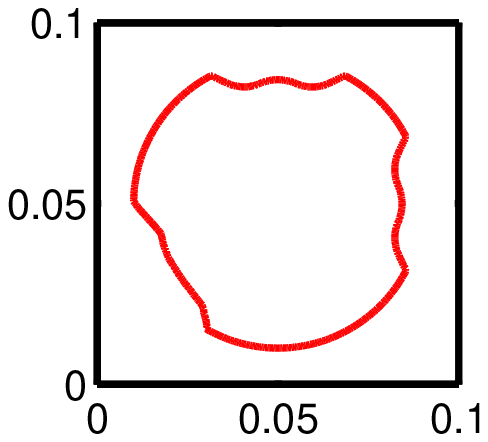}}\\
   \subfigure[ $t = 15$]{\includegraphics[height=1.5in]{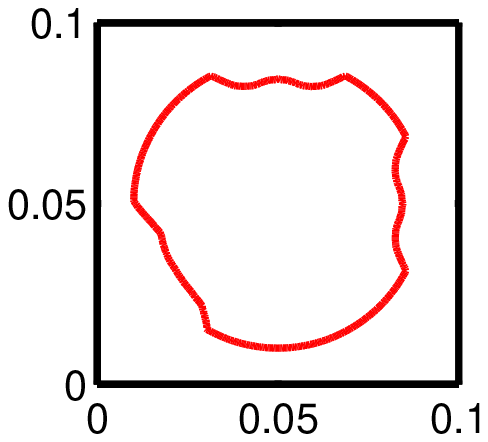}}
   \subfigure[ $t = 20$]{\includegraphics[height=1.5in]{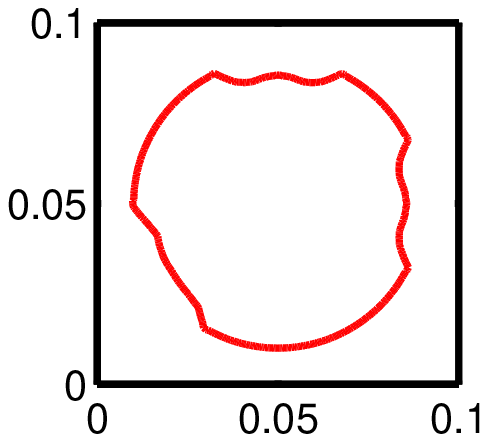}\label{if-e}}
   \subfigure[$t = 25$ days]{\includegraphics[height=1.5in]{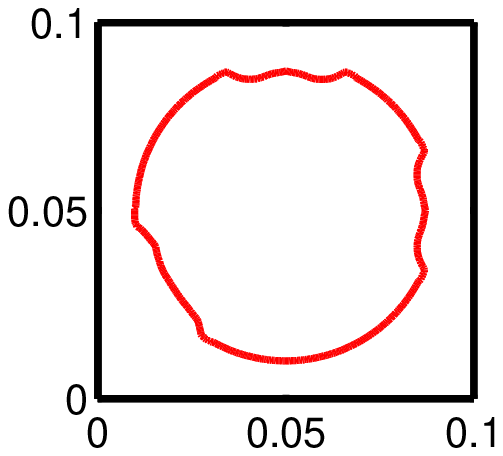}\label{if-f}} \\
   \subfigure[ $t = 30$]{\includegraphics[height=1.5in]{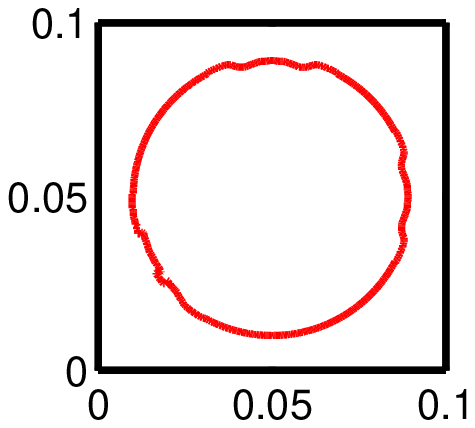}\label{if-g}}
   \subfigure[ $t = 35$]{\includegraphics[height=1.5in]{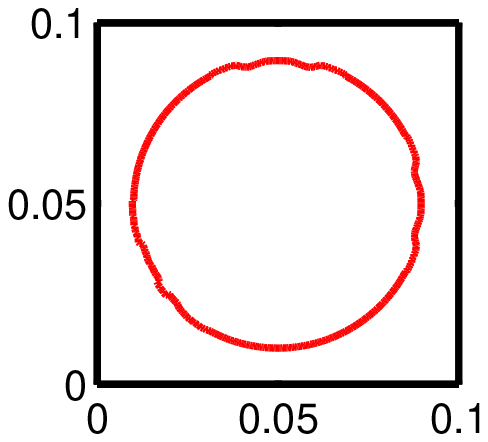}}
   \subfigure[$t = 150$ days]{\includegraphics[height=1.5in]{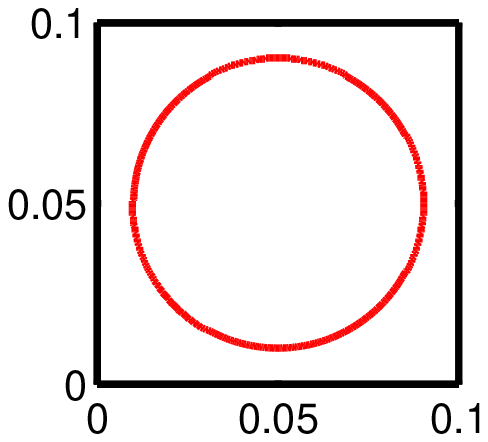}\label{if-i}}
   \caption{{\bf{The evolution of the interface between bone and marrow regions due to remodeling at three distinct remodeling cites.}} The speed in the normal direction is given by the change in the bone mass (\ref{eq:normv}). Figure \ref{if-a} shows the initial interface before any remodeling has begun. The interior of the circle represents the bone region while the exterior is marrow. The interface between bone and marrow moves in and out, i.e. in the normal direction, due to changes in the populations of osteoclasts and osteoblasts.}
   \label{interface}
 \end{figure}

 We begin remodeling at three distinct sites by perturbing the osteoclast population away from steady state, figure \ref{oc-a}. This initiates bone resorption. The effects of remodeling on the interface after 5 days of bone resorption is shown in figure \ref{if-b}. Note that the interface moves inward (toward the interior region on bone) due resorption since there is an increase in osteoclast population away from steady state. The osteoclast population quickly returns to steady state, figure \ref{oc-f}. In accordance with equation \ref{eq:bonemass1} when the osteoclast population returns to steady state the motion should cease to be inward. This can be seen in figures \ref{if-e} and \ref{if-f}.  In the meantime the osteoblast population increases from steady state, figures \ref{ob-b}, \ref{ob-c}, and new bone is being formed. The effect of remodeling on the interface during the early stages of formation of new bone by osteoblasts is shown if figures \ref{if-f} and \ref{if-g} where the indentions become filled in. After reaching approximately maximum population density, figures \ref{ob-e}, \ref{ob-f}, the osteoblasts return to steady state by which time bone has been fully remodeled, figures \ref{ob-i}, \ref{if-i}.

  Comparing the results in the three figures \ref{osteoclasts},\ref{osteoblasts} and \ref{interface} at corresponding times shows how the changes in the cell populations are manifested as physical changes in the bone surfaces at three different remodeling sites. The interface moves in the normal direction as a result of remodeling in accord with (\ref{eq:sysone}). While here we have initiated remodeling at each of the three sites at the same instant this is not a requirement of this modeling approach. Remodeling at different sites can easily be taken to be out of phase.

\begin{figure}[!ht]
  \centering
  \includegraphics[width=60mm,height=60mm]{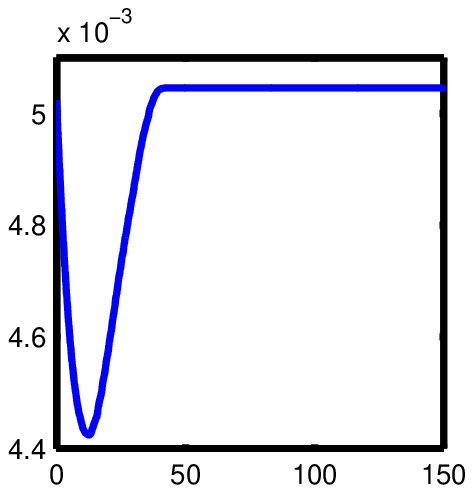}
  \caption{{\bf{The amount of change due to remodeling.}} The units are days for the horizontal axis and $\mathrm{mm}^{2}$ for the vertical axis. We arrived at this by computing the interior area at each time in the evolution of the interface. This can be compared with figure \ref{komFig}. The differences in geometry (i.e. velocity times radius squared here, and local bone mass in \ref{komFig}) results in a slightly faster, but qualitatively similar, cycle than in figure \ref{komFig}. }\label{area1}
\end{figure}

 We have described in figure \ref{area1} quantitatively the change in the bone by plotting the area (in square microns) interior to bone, this can be compared with figure 2, which shows the bone mass as a function of time. We remark that bone mass resorption and formation occurs somewhat faster in figure \ref{area1} as we add more spatial dimensions due to the geometry, (i.e. proportional to radius squared versus proportional to local thickness). We note however, the qualitative agreement with figure 2 in \cite{komarova2003}.

 As discussed previously, in the work \cite{komarova2010}, a mathematical model including explicit spatial terms is considered. To illustrate the
 flexibility of the level-set approach we discuss how this can be done here as well. The first step is the inclusion of signaling pathways not through exponents in a power law approximation but explicitly. For example the RANKL/OPG pathway can be represented explicitly by modifying (\ref{eq:clasts1}) to include the effect of RANKL on osteoclast arriving at
  \begin{subequations}
   \begin{align}
     \frac{dC}{dt} & = \ao C^{g_{11}}- \bo C + \kappa_{1}\frac{u}{\lambda + u}H(\max\{0,C - \bar{C}\})C. \label{eq:clasts2}
   \end{align}
 Then the RANKL/OPG field is represented by equations for the concentrations of each chemical species. These are
  \begin{align}
     u_{t} & = D(x) \Delta u - \kappa_{2} \frac{u}{\lambda + u}H(\max\{0,C - \bar{C}\})C \nonumber \\
      & + a_{R}\max\{0,B - \bar{B}\} - \kappa_{3}uv, \label{eq:rankl} \\
   v_{t} & = D(x) \Delta v  + a_{O}\max\{0,B - \bar{B}\} - \kappa_{3}uv, \label{eq:opg}
 \end{align}
 where $u$ denotes the concentration of RANKL and $v$ the concentration of OPG.
 \end{subequations}
 The equations for the osteoblast population and change in bone mass remain as in (\ref{eq:blasts1}) and (\ref{eq:normv}) respectively. Note that
  while in \cite{komarova2010} steering is controlled by the chemotaxis of osteoclasts toward RANKL concentrations, here it is controlled by the geometry through the level-set function as above. Our computational results including explicit spatial terms for the chemical species showed an evolution of the interface similar to figure \ref{interface}.

%\subsection*{Subsection 2}

Based on the local dynamics of bone remodeling as developed in \cite{komarova2003,komarova2005,ryser1,komarova2010} we have developed a new approach to modeling the spatial effects of bone remodeling. Through the use of the level-set equation we link local dynamics with geometric effects and spatial behavior such as steering of a BMU in a way that is appropriate on many different spatial scales. The approach taken is useful but not limited to describing the effects of bone remodeling in complex geometries such as is seen in the remodeling of trabecular bone. The extension of this model
 to various other situations is straightforward and software for implementing the level-set equation approach numerically is widely available (see e.g. \cite{michell1} and
 references therein).

 The development of mathematical and computational models of biological phenomena is valuable in design and analysis of experiments and aids in building intuition and conceptual understanding of biological processes. These models can also be useful to the biomedical community. Analysis
 and simulation based on mathematical and computational models can suggest how and where pathological behavior occurs. Examples of this in bone remodeling can be seen in \cite{komarova2003,komarova2005,ayati2010} and the modeling approach in this paper can be used in the same way by including the appropriate modifications. Modeling can also aid in considering treatments and therapies for osteopathies (see e.g. \cite{akchurin2008}).  We feel the type of modeling presented in this work can be especially helpful in ruling out ineffective treatments and therapies.

 For future work more sophisticated numerical schemes for solving the level-set equation and computing solutions of the model equations will be employed.
 In particular since remodeling occurs primarily near a small region of the interface, a local level-set method such as in \cite{peng1999,salac2008}  can be used to speed up computations. It may also be appropriate in the future to include further details of the cytokine signaling, something that is likely to involve coupling additional differential equations with the existing model. This could call for a need of more efficient and accurate numerical algorithms to carry out simulations, particularly important in moving to three dimensional problems. While the level-set equation scales well in higher dimensions, numerical solutions of other differential equations become much more expensive from a computational standpoint.
 We feel the level-set approach taken here will provide a supplement or alternative to other models of the fundamental process of bone remodeling. Further development in this direction will allow for applications to a variety of biomedical issued related to bone remodeling at a variety of scales.

 \section*{Acknowledgments}

This work was supported by the National Science Foundation grant number DMS-0914514 and by the National Institutes of Health Challenge
  grant number 1 RC1 AR058403-01.

\bibliography{remodelingBib}

%\medskip
%%% The received and accepted dates
%Received September 10, 2006; Accepted February 10, 2007.

\medskip

\end{document}